\RequirePackage{fix-cm}
\documentclass[smallextended]{svjour3}       
\smartqed  
\usepackage{graphicx}

\usepackage{amsmath}
\usepackage{amssymb}
\usepackage{bm}
\usepackage{braket}
\begin{document}

\title{Strong coupling effects on specific heat in the BCS-BEC crossover
}

\author{Daisuke Inotani \and Pieter van Wyk \and Yoji Ohashi
}


\institute{D. Inotani \at
              Department of Physics, Keio University, 3-14-1 Hiyoshi, Kohoku-ku, Yokohama 223-8522,
Japan. \\
              Tel.: +81-45-566-1454\\
              Fax: +81-45-566-1672\\
              \email{dinotani@rk.phys.keio.ac.jp}           
}

\date{Received: \today / Accepted: \today}

\maketitle

\begin{abstract}
We theoretically investigate strong-coupling effects on specific heat at constant volume $C_{\rm V}$ in a superfluid Fermi gas with a tunable interaction associated with Feshbach resonance. Including fluctuations of the superfluid order parameter within the strong-coupling theory developed by Nozi\`eres and Schmitt-Rink, we calculate the temperature dependence of $C_{\rm V}$ at the unitarity limit in the superfluid phase. We show that, in the low temperature region, $T^3$-behavior is shown in the temperature dependence of $C_{\rm V}$. This result indicates that the low-lying excitations are dominated by the gapless Goldstone mode, associated with the phase fluctuations of the superfluid order parameter. Since the Goldstone mode is one of the most fundamental phenomena in the Fermionic superfluidity, our results are useful for further understanding how the pairing fluctuations affects physical properties in the BCS-BEC crossover physics below the superfluid transition temperature. 
\end{abstract}
\section{Introduction}
\label{intro}
Since the BCS-BEC crossover, where the superfluid properties continuously change from the weak-coupling BCS (Bardeen-Cooper-Schrieffer)-type to the BEC (Bose-Einstein condensation) of tightly binding molecular Bose gas as increasing the interaction strength, has been realized in $^{40}$K\cite{Regal} and $^{6}$Li\cite{Zwierlein2004,Kinast2004,Bartenstein2004} Fermi gases, strong-coupling effects on various physical quantities in this system, such as single-particle properties\cite{Tsuchiya2009,Tsuchiya2011}, spin susceptibility\cite{Tajima2014}, and thermodynamic properties\cite{Haussmann,Hu2006pra,Pieter2016}, have extensively investigated. Recently, in ultracold Fermi gas, the great progress of experimental technique makes us enable to measure various thermodynamic quantities\cite{Kinast2005,Luo2009,Horikochi2010,Nascimbene2010,Ku}.
\par
In the BCS-BEC crossover phenomena below the superfluid transition temperature, the low-energy excitations give us a useful information for further understanding this phenomenon. While in the weak-coupling BCS regime there are two kinds of fundamental excitations, i.e. the Fermionic single-particle excitations associated with pair breaking and the gapless Goldstone mode associated with the phase fluctuations of the superfluid order parameter $\Delta$, in the strong-coupling BEC regime the former is strongly suppressed due to the large binding energy of the pairs. Although in the low temperature region where $T \ll \Delta$ the low-lying excitations are always dominated by the Goldstone mode in the entire interaction regime, since the sound velocity of this mode strongly depends on the interaction strength\cite{Ohashi2003}, a characteristic temperature where the system starts to be dominated by the Goldstone mode can be used to determine the region where the phase fluctuations become remarkable.
\par
For this purpose, we focus on the specific heat at constant volume $C_{\rm V}$, which has been known to be sensitive to pairing fluctuations above the superfluid transition temperature $T_{\rm c}$\cite{Pieter2016}. As well known, while within the weak-coupling mean-field theory in which the gapless Goldstone mode is ignored, $C_{\rm V}$ is dominated by the pair breaking leading $e^{-\Delta/T}$-behavior in the low temperature region, in the ideal Bose gas, $C_{\rm V}$ shows $T^{3/2}$-behavior. On the other hand, when the thermodynamic properties are dominated by the gapless Goldstone mode, $C_{\rm V}$ should be proportional to $T^{3}$\cite{Haussmann}. Thus, the temperature dependence of $C_{\rm V}$ is sensitive to what kind of excitations is remarkable. In this sense, from the detailed temperature dependence of $C_{\rm V}$, we might determine the region where the gapless Goldstone mode associated with the phase fluctuations dominates the thermodynamic properties.
\par
In this paper, we theoretically calculate $C_{\rm V}$ at the unitarity limit below the superfluid transition temperature $T_{\rm c}$ within a strong-coupling NSR theory developed by Nozi\`eres and Schmitt-Rink\cite{Hu2006pra,Ohashi2003,NSR,Fukushima2007,Hu2006epl}, in which the contributions from the gapless Goldstone mode are taken into account. We find that in the low temperature region where the thermal transfer from the gapless Goldstone mode to the single-particle excitations is sufficiently suppressed, $C_{\rm V}$ shows $T^{3}$-behavior. Throughout this paper, we take $\hbar=k_{\rm B}=1$, and the system volume $V$ is taken to be unity, for simplicity.
\section{Formulation}
We consider a two-component Fermi gas with a contact-type attractive interaction associated with Feshbach resonance, described by the Hamiltonian
\begin{equation}
H=\sum_{\bm p,\sigma} \xi_{\bm p}c_{\bm p,\sigma}^{\dagger}c_{\bm p,\sigma}
-U\sum_{{\bm p},{\bm p}',{\bm q}} 
c_{{\bm p}+{\bm q}/2,\uparrow}^\dagger c_{-{\bm p}+{\bm q}/2,\downarrow}^\dagger
c_{-{\bm p}'+{\bm q}/2,\downarrow}c_{{\bm p}'+{\bm q}/2,\uparrow}.
\label{eq1}
\end{equation}
Here $c_{\bm p,\sigma}$ denotes an annihilation operator of a Fermi atom with the momentum ${\bm p}$ and the pseudospin $\sigma=\uparrow, \downarrow$, and $\xi_{\bm p}=p^2/(2m) -\mu$ is the kinetic energy measured from the chemical potential $\mu$ (where $m$ is the atomic mass). The second term in the Hamiltonian Eq. (\ref{eq1}) describes the contact-type attractive interaction with a tunable coupling constant $U>0$, which is conveniently measured in terms of the observable $s$-wave scattering length as 
\begin{equation}
{4\pi a_s \over m}=
\frac{-U}
{1-U\sum_{{\bm p}}^{p_{\rm c}}
\frac{1}{2\varepsilon_{\bm p}}}
.
\label{eq2}
\end{equation}
Here, $p_{\rm c}$ is a cutoff momentum. Using this scale, the weak- and strong-coupling region are characterized by the region where $(k_{\rm F} a_s)^{-1} < -1$ and $(k_{\rm F} a_s)^{-1} > 1$, respectively. The region between them ($-1<(k_{\rm F} a_s)^{-1} < 1$) is usually referred as the crossover region, in which pairing fluctuations might be remarkable. 
\par
To investigate the superfluid properties of this system, as usual, it is convenient to write the Hamiltonian Eq. (\ref{eq1}) in the Nambu representation\cite{Ohashi2003} by introducing the superfluid order parameter $\Delta=-U\sum_{\bm p} \left\langle c_{-{\bm p},\downarrow}c_{{\bm p},\uparrow} \right\rangle$, as
\begin{equation}
H=\sum_{\bm p} \psi_{\bm p}^{\dagger} 
\left( \xi_{\bm p}\tau_3  - \Delta \tau_1 \right)
\psi_{\bm p}
-U\sum_{\bm q} \rho_+\left( {\bm q} \right) \rho_-\left( -{\bm q} \right).
\label{eq3}
\end{equation}
Here $\psi_{\bm p}= \left( c_{{\bm p},\uparrow}  c_{-{\bm p},\downarrow}^{\dagger}  \right)^T$
is the two-component Nambu spinor operator, $\tau_i$ ($i=1,2,3$) is Pauli matrices acting on the particle-hole space, and
\begin{equation}
\rho_{\pm} \left({\bm q }\right)= \sum_{\bm {p}} \psi_{{\bm p}+\frac{\bm q}{2}}^{\dagger} \tau_{\pm} \psi_{{\bm p}-\frac{\bm q}{2}}
\label{eq5}
\end{equation}
is the generalized density operator describing fluctuations of $\Delta$, where
$\tau_{\pm}=\frac{1}{2} \left(\tau_1 \pm i \tau_2 \right)$. In Eq. (\ref{eq3}), $\Delta$ is taken to be real without loss of generality.
\par
\begin{figure}
  \centering
  \includegraphics[width=1\textwidth]{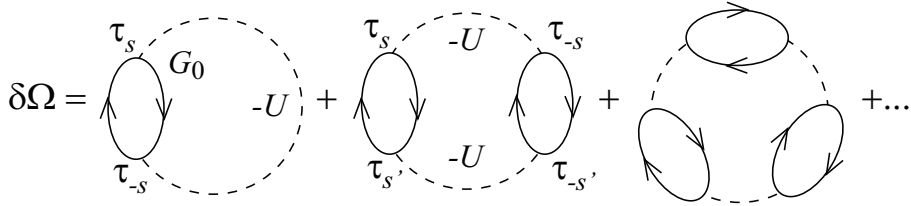}
\caption{Strong-coupling corrections to the thermodynamic potential within NSR theory. The solid and dashed line represent the mean-field single-particle Green's function $G_0$ and the contact-type attractive interaction $-U$, respectively. $\tau_{s=\pm}= (\tau_1+i\tau_2)/2$.
}
\label{fig1}      
\end{figure}
The strong-coupling corrections from the second term in Eq. (\ref{eq3}) to the thermodynamic properties are taken into account within NSR theory\cite{Hu2006pra,Ohashi2003,NSR,Fukushima2007,Hu2006epl}, in which the thermodynamic potential is given as the sum
$\Omega=\Omega_{\rm {MF}}+\delta\Omega$ of the mean-field part $\Omega_{\rm MF}$ and the strong-coupling corrections $\delta \Omega$. The mean field part is given by
\begin{equation}
\Omega_{\rm {MF}}=
-\frac{m\Delta^2}{4\pi a_s}
+\sum_{\bm p} \left(  \xi_{\bm p} - E_{\bm p} + \frac{\Delta^2}{2\varepsilon_{\bm p}}
 \right)
-2T\sum_{\bm p} \ln \left( 1+e^{-\frac{E_{\bm p}}{T}}\right),
\label{eq7}
\end{equation}
where $E_{\bm p}=\sqrt{\xi_{\bm p}^2 + |\Delta|^2}$ is the Bogoliubov single-particle excitation spectrum. The strong-coupling corrections to the thermodynamic potential  $\delta \Omega$ are described by the Feynman diagrams shown in Fig. \ref{fig1}, which give
\begin{equation}
\delta\Omega=-\frac{1}{2\beta} \sum_{{\bm q}, i\nu_n}\mathrm{Tr} \left[ \ln \hat{\Gamma}({\bm q}, i\nu_n)  \right].
\label{eq8}
\end{equation}
Here, $\hat{\Gamma}({\bm q}, i\nu_n)=-U/[1+U\hat{\Pi}({\bm q},i\nu_n)]$ is the 2$\times$2 particle-particle scattering matrix, where
\begin{align}
\hat{\Pi}({\bm q},i\nu_n)&=
\left(
\begin{array}{cc}
\Pi_{-+}({\bm q}, i\nu_n) & \Pi_{--}({\bm q}, i\nu_n) \\
\Pi_{++}({\bm q}, i\nu_n) & \Pi_{+-}({\bm q}, i\nu_n) \\
\end{array}
\right),
\label{eq9}
\\
\Pi_{ss'}({\bm q}, i\nu_n)&=\frac{1}{\beta}\sum_p 
\mathrm{Tr} \left[ \tau_{s} G_0 \left( {\bm p+\frac{q}{2}},i\omega_n \right)
\tau_{s'} 
G_0\left( {\bm p-\frac{q}{2}}, i\omega_n-i\nu_n \right)
\right],
\label{eq10}
\end{align}
is the lowest-order 2$\times$2 matrix pair correlation function describing the fluctuation of the superfluid order parameter, and
\begin{equation}
G_{0} \left({\bm p} ,i\omega_n \right)=
\frac{1}{i\omega_{n}-\xi_{\bm p} \tau_3 +\Delta \tau_1}
\label{eq11}
\end{equation}
is the 2$\times$2 matrix single-particle Green's function in the mean-field level. 
\par
The specific heat at constant volume $C_{\rm V}$ is obtained from a thermodynamic relation
\begin{equation}
C_{\rm V}=\left(\frac{\partial E}{\partial T}
\right)_{N}.
\label{eq12}
\end{equation}
Here $E$ is the internal energy, which is given by the Legendre transformation from $\Omega$ as  
\begin{equation}
E=\Omega+TS+\mu N
=\Omega-T
\left(\frac{\partial \Omega}{\partial T}
\right)_{\mu}
-\mu\left( 
\frac{\partial \Omega}{\partial \mu}
\right)_{T}.
\label{eq13}
\end{equation}
In this paper, by numerically carrying out the derivative of $E$ with respect to the temperature $T$ in Eq. (\ref{eq12}), $C_{\rm V}$ is calculated. In this procedure, we first determine the chemical potential $\mu$ and the superfluid order parameter $\Delta$ as a function of $T$ by self-consistently solving  the gap equation, which is obtained from the Thouless criterion ${\rm det} \left[\Gamma({\bm q} = 0, i\nu_n = 0)^{-1} \right] = 0$ as
\begin{equation}
1=\frac{4\pi a_s}{m} \sum_{\bf p} \left( \frac{1}{2E_{\bf p}} \tanh \frac{2E_{\bf p}}{2T} 
- \frac{1}{2\varepsilon_{\bf p}} \right),
\label{eq14}
\end{equation}
together with the particle number equation 
\begin{align}
N&=-\left( 
\frac{\partial \Omega}{\partial \mu}
\right)_{T}
=-\left( 
\frac{\partial \Omega}{\partial \mu}
\right)_{T,\Delta}
-\left( 
\frac{\partial \Omega}{\partial \Delta}
\right)_{T,\mu}
\left( 
\frac{\partial \Delta}{\partial \mu}
\right)_{T}.
\label{eq15}
\end{align}
We note that the Thouless criterion guarantees the existence of the gapless Goldstone mode, of which the dispersion relation is obtained by solving the equation ${\rm{det}}\left[ \Gamma \left({\bm q}, i\nu_n \to \omega+i\delta \right)^{-1} \right]=0$. In the low temperature limit ($T \ll \left\| \mu \right\|, \Delta $), indeed, we obtain the well-known Goldstone mode $\omega_q=v_\phi q$ in the low energy region\cite{Ohashi2003}, with the sound velocity
\begin{align}
v_\phi&=\frac{\Delta^2}{2m}
\left(\sum_{\bm p} \frac{1}{E_p^3}\right)
\frac{\sum_{\bm p} \frac{\xi_p}{E_p^3}+2\Delta^2\sum_{\bm p} \frac{\varepsilon_p}{E_p^5}}
{\Delta^2 \left(\sum_{\bm p} \frac{1}{E_p^3}\right)^2+\left(\sum_{\bm p} \frac{\xi_p}{E_p^3}\right)^2}
,
\label{eqvs}
\end{align}
which is expected to give a $T^3$ contribution to the specific heat. We also briefly mention that the superfluid transition temperature $T_{\rm c}$ is obtained by solving Eqs. (\ref{eq14}) and (\ref{eq15}) with $\Delta=0$, and $\mu$ above $T_{\rm c}$ is simply given by solving only Eq. (\ref{eq15}) with $\Delta=0$. In the next section, in addition to the results in the superfluid phase, we also show the results in the normal phase\cite{Pieter2016}, for comparison. 
\section{Result}
\begin{figure}
  \centering
  \includegraphics[width=1\textwidth]{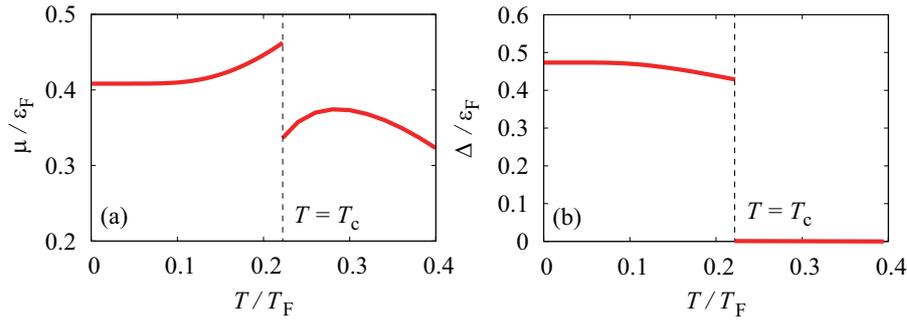}
\caption{Calculated (a) chemical potential $\mu$ and (b) superfluid order parameter $\Delta$ of two-component Fermi gas with an attractive interaction as a function of temperature at the unitarity limit $(k_{\rm F}a_s)^{-1}=0$. Within NSR theory, the superfluid transition temperature $T_{\rm c}= 0.222T_{\rm F}$.
}
\label{fig2}      
\end{figure}
Figure \ref{fig2} shows the temperature dependence the chemical potential $\mu$, as well as the superfluid order parameter $\Delta$ in the unitarity limit, determined from Eqs. (\ref{eq14}) and (\ref{eq15}). Here, we note that $\Delta$ in Fig. \ref{fig2} has finite value even at $T=T_{\rm c}$. However, it is known that this behavior indicating a first-order phase transition is an artifact in our approximation. Same problem has been reported in other diagrammatic strong-coupling theories, such as non-self-consistent $T$-matrix theory\cite{TMA}, as well as self-consistent $T$-matrix theory\cite{SCTMA}. Since this unphysical behavior of $\Delta$ might affect our results just below $T_{\rm c}$, we will focus on the low-temperature behavior of $C_{\rm V}$, and we leave this problem as a future work. We also note that the obtained $\Delta(T)$ is always larger than $T$ below $T_{\rm c}$.
Thus, in the superfluid phase at the unitarity limit, the single-particle excitations associated with the pair-breaking, which is characterized by an energy scale $2\Delta$, are thermally suppressed, and the thermodynamics of this system is expected to be dominated by the gapless Goldstone mode, as well as the thermal transfer from the collective mode to the single-particle excitations.  
\par
\begin{figure}
  \centering
  \includegraphics[width=1\textwidth]{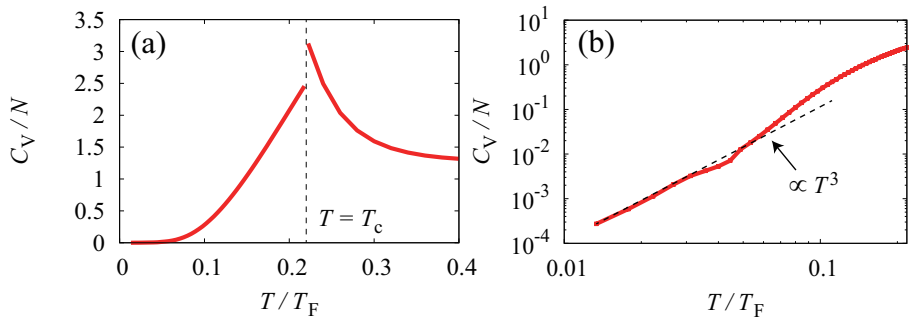}
\caption{(a) Calculated specific heat at constant volume $C_{\rm V}$ of two-component Fermi gas with an attractive interaction at the unitarity limit. (b) Results below the superfluid transition temperature $T_{\rm c}= 0.222T_{\rm F}$ in log-scale. In panel (b), a dashed line parallel to $T^3$ is also shown.  
}
\label{fig3}      
\end{figure}
Figure \ref{fig3} shows calculated temperature dependence of $C_{\rm V}$ at the unitarity limit. Starting from the high temperature region, in normal phase $C_V$ is found to increase as approaching $T_{\rm c}$, that is in contrast to one of the ideal Fermi gas where $C_{\rm V}$ monotonically decreases as decreasing $T$. As discussed in our previous paper\cite{Pieter2016}, this enhancement of $C_{\rm V}$ near $T_{\rm c}$ originates from the appearance of the preformed Cooper pairs associated with a strong-attractive interaction. We also find that a jump of $C_{\rm V}$ across $T_{\rm c}=0.22T_{\rm F}$, as shown in the ordinary BCS superconductors. However, as mentioned above, since our results just below $T_{\rm c}$ are not reliable due to the artificial first-order phase transition, we do not further discuss this point. 
\par
In the superfluid phase below $T_{\rm c}$, $C_{\rm V}$ monotonically decreases as decreasing $T$, and eventually vanishes at $T=0$, as expected. As shown in Fig \ref{fig3} (b), in the low temperature region ($T \lesssim 0.03T_{\rm F}$), we find that $C_{\rm V}$ is proportional to $T^3$, that indicates that $C_{\rm V}$ is dominated by the gapless Goldstone mode. 
\begin{figure}
  \centering
  \includegraphics[width=0.5\textwidth]{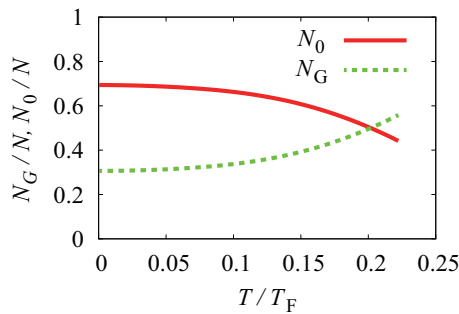}
\caption{Temperature dependence of the number of unpaired atom and condensed pairs $N_0$ and the contribution to the particle number from the gapless Goldstone mode $N_{\rm G}$. 
}
\label{fig4}      
\end{figure}
To more clearly see this, we conveniently write the particle number Eq. (\ref{eq15}) as the sum $N=N_0+N_{\rm G}$ of the contribution from the Goldstone mode
\begin{align}
N_{\rm G}=-\left( 
\frac{\partial \Omega}{\partial \mu}
\right)_{T,\Delta}
\label{eq16}
\end{align}
and the others $N_0$. Here, we mention that, although $N_{\rm G}$ includes the contribution from both the phase and the amplitude fluctuations of the superfluid order parameter, because the amplitude fluctuations are rapidly suppressed as developing $\Delta$, at least, in the low-temperature region $T \ll \Delta$, $N_{\rm G}$ can be regarded as the contribution from the Goldstone mode. Noting that the single-particle excitations associated with pair breaking can also be ignored in this low temperature region, the low-temperature behavior of $C_{\rm V}$ originates from the gapless Goldstone mode, as well as the thermal transfer from the Goldstone mode to the single-particle contributions. As shown in Fig. \ref{fig4}, as decreasing $T$, the latter effects are gradually suppressed, and when $T \lesssim 0.05T_{\rm F}$, $N_{\rm G}$ becomes almost constant. Then, $T^3$-dependence coming from the Goldstone mode becomes dominant in $C_{\rm V}$. 
\section{Conclusion}
\label{summary}
In this paper, we have theoretically investigated the effects of fluctuations of the superfluid order parameter on the specific heat at constant volume $C_{\rm V}$ at the unitarity limit within a strong-coupling NSR theory. We found that in the low-temperature region, where the thermal transfer from the gapless Goldstone mode to the single-particle excitations is sufficiently suppressed, $C_{\rm V}$ exhibits a $T^3$-dependence. Since the $T^3$-dependence comes from the gapless Goldstone mode associated with the phase fluctuations of the superfluid order parameter, our results indicate that the temperature dependence of $C_{\rm V}$ might be useful to determine the region where the phase fluctuations dominate the thermodynamic properties. Furthermore, since the gapless Goldstone mode always exists in the superfluid phase, but its properties remarkably depend on the interaction strength, it is our future problem how the temperature dependence of $C_{\rm V}$ changes as varying the interaction strength.
\begin{acknowledgements}
This work was supported by KiPAS project in Keio University. DI was supported by Grant-in-aid for Scientific Research from JSPS in Japan (No.JP16K17773). YO was supported by Grant-in-aid for Scientific Research from MEXT and JSPS in Japan (No.JP18K11345, No.JP18H05406, No.JP16K05503).  
\end{acknowledgements}

\begin{thebibliography}{100}
\bibitem{Regal} C. A. Regal, M. Greiner, and D. S. Jin, Phys. Rev. Lett. {\bf 92}, 040403 (2004).
\bibitem{Zwierlein2004} M. W. Zwierlein, C. A. Stan, C. H. Schunck, S. M. F. Raupach, A. J. Kerman, and W.
Ketterle, Phys. Rev. Lett {\bf 92}, 120403 (2004).
\bibitem{Kinast2004} J. Kinast, S. L. Hemmer, M. E. Gehm, A. Turlapov, and J. E. Thomas, Phys. Rev. Lett. {\bf 92},
150402 (2004).
\bibitem{Bartenstein2004} M. Bartenstein, A. Altmeyer, S. Riedl, S. Jochim, C. Chin, J. H. Denschlag, and R. Grimm,
Phys. Rev. Lett 92, 203201 (2004).
\bibitem{Tsuchiya2009} S. Tsuchiya, R. Watanabe, and Y. Ohashi, Phys. Rev. A {\bf 80}, 033613 (2009).
\bibitem{Tsuchiya2011} S. Tsuchiya, R. Watanabe, and Y. Ohashi, Phys. Rev. A {\bf 84}, 043647 (2011).
\bibitem{Tajima2014} H. Tajima, T. Kashimura, R. Hanai, R. Watanabe, and Y. Ohashi, Phys. Rev. A {\bf 89}, 033617 (2014).
\bibitem{Haussmann} R. Haussmann, W. Rantner, S. Cerrito, and W. Zwerger, Phys. Rev. A, {\bf 75}, 023610 (2007). 
\bibitem{Hu2006pra} H. Hu, X.-J. Liu, and P. D. Drummond, Phys. Rev. A {\bf 73}, 023617 (2006).
\bibitem{Pieter2016} P. van Wyk, H. Tajima, R. Hanai, Y. Ohashi, Phys. Rev. A {\bf 93}, 013621 (2016).
\bibitem{Kinast2005} J. Kinast, A. Turpalov, J. E. Thomas, Q. Chen, J. Stajic and K. Levin, Science, 307, 1296 (2005).
\bibitem{Luo2009} L. Luo and J. E. Thomas, J. Low. Temp. Phys 154, 1 (2009).
\bibitem{Horikochi2010} M. Horikoshi, S. Nakajima, M. Ueda and T. Mukaiyama, Science 327, 442 (2010).
\bibitem{Nascimbene2010} S. Nascimbene, N. Navon, K. J. Jiang, F. Chevy and C. Salomon, Nature 463, 1057 (2010).
\bibitem{Ku} M. J. H. Ku, A. T. Sommer, L. W. Cheuk, and M. W. Zwierlein, Science. {\bf 335}, 563-567 (2012).
\bibitem{Ohashi2003} Y. Ohashi and A. Griffin, Phys. Rev. A {\bf 67}, 063612 (2003).
\bibitem{NSR} P. Nozi\`eres, and S. Schmitt-Rink, J. Low Temp. Phys. {\bf 59}, 195 (1985).
\bibitem{Fukushima2007} N. Fukushima, Y. Ohashi, E. Taylor and A. Griffin, Phys. Rev. A {\bf 75}, 033609 (2007).
\bibitem{Hu2006epl} H. Hu, X.-J. Liu, and P. D. Drummond, Europhys. Lett. {\bf 74}, 574 (2006).
\bibitem{TMA} R. Watanabe, S. Tsuchiya, Y. Ohashi, Phys. Rev. A {\bf 82}, 043630 (2010)
\bibitem{SCTMA} R. Haussmann, W. Rantner, S. Cerrito and W. Zwerger, Phys. Rev. A {\bf 75}, 023610 (2007).
\end{thebibliography}


\end{document}